\pdfoutput=1
\documentclass[twocolumn,english,aps,prl,twolocumns]{revtex4}
\usepackage[T1]{fontenc}
\usepackage[latin9]{inputenc}
\usepackage[letterpaper]{geometry}
\usepackage{color}
\usepackage{units}
\usepackage{textcomp}
\usepackage{amsthm}
\usepackage{amsmath}
\usepackage{graphicx}
\usepackage{amssymb}

\makeatletter
\@ifundefined{textcolor}{}
{%
 \definecolor{BLACK}{gray}{0}
 \definecolor{WHITE}{gray}{1}
 \definecolor{RED}{rgb}{1,0,0}
 \definecolor{GREEN}{rgb}{0,1,0}
 \definecolor{BLUE}{rgb}{0,0,1}
 \definecolor{CYAN}{cmyk}{1,0,0,0}
 \definecolor{MAGENTA}{cmyk}{0,1,0,0}
 \definecolor{YELLOW}{cmyk}{0,0,1,0}
 }
\theoremstyle{plain}

\makeatother

\usepackage{babel}

\begin{document}

\title{Lifetime measurements ($T_{1}$) of electron spins in Si/SiGe quantum
dots}

\author{Robert~R.~Hayes }

\email{rrhayes@HRL.com}

\author{Andrey~A.~Kiselev}

\author{Matthew G. Borselli}

\author{Steven S. Bui}

\author{Edward~T.~Croke~III }

\author{Peter W. Deelman}

\author{Brett M. Maune}

\author{Ivan Milosavljevic}

\author{Jeong-Sun Moon}

\author{Richard S. Ross}

\author{Adele E. Schmitz}

\author{Mark F. Gyure}

\author{Andrew T. Hunter}

\affiliation{HRL Laboratories, LLC, Malibu, CA 90265, USA}

\date{\today}
\begin{abstract}
We have observed the Zeeman-split excited state of a spin-$\nicefrac{1}{2}$
multi-electron Si/SiGe depletion quantum dot and measured its spin
relaxation time $T_{1}$ in magnetic fields up to $2$~T. Using a
new step-and-reach technique, we have \textcolor{black}{experimentally
verified }the $g$-value of $2.0\pm0.1$ for the observed Zeeman doublet.
We have also measured $T_{1}$ of single- and multi-electron spins
in InGaAs quantum dots. The lifetimes of the Si/SiGe system are appreciably
longer than those for InGaAs dots for comparable magnetic field strengths,
but both approach one second at sufficiently low fields ($<1$~T
for Si, and $<0.2$~T for InGaAs). 
\end{abstract}
\maketitle
\textbf{Introduction }Gate-defined quantum dots (QD) in the Si/SiGe
material system have been touted as one of the more promising candidates
for spin-based quantum computation, primarily because of the long
decoherence time $T_{2}$ expected for electron spins in Si \cite{Ericksson}.
Because $T_{1}$ establishes a natural upper bound on $T_{2}$, a
measurement of the spin relaxation time $T_{1}$ for isolated electrons
in any Si-based system has been an important and actively-pursued
objective \cite{Si-based,Simmons}. Although there has been significant
progress in fabricating and testing few-electron Si/SiGe dots \cite{Simmons},
there has been, up until now, no direct measurement showing the anticipated
long spin-relaxation lifetimes. Indeed, there has been no direct measurement
of a Zeeman splitting for either single- or few-electron states in
any Si-based quantum dot.

In this Letter, we report the first direct experimental confirmation
of a Zeeman spin excited state in a few-electron Si/SiGe quantum dot,
and a measurement of its lifetime as a function of magnetic field.
We also report comparable data for InGaAs dots, used as a testbed
for our measurement techniques. These dots could be emptied, thus
allowing us to measure single-electron spins. 

\begin{figure}
\noindent \begin{centering}
\includegraphics[bb=120bp 195bp 580bp 440bp,clip,width=3.2in]{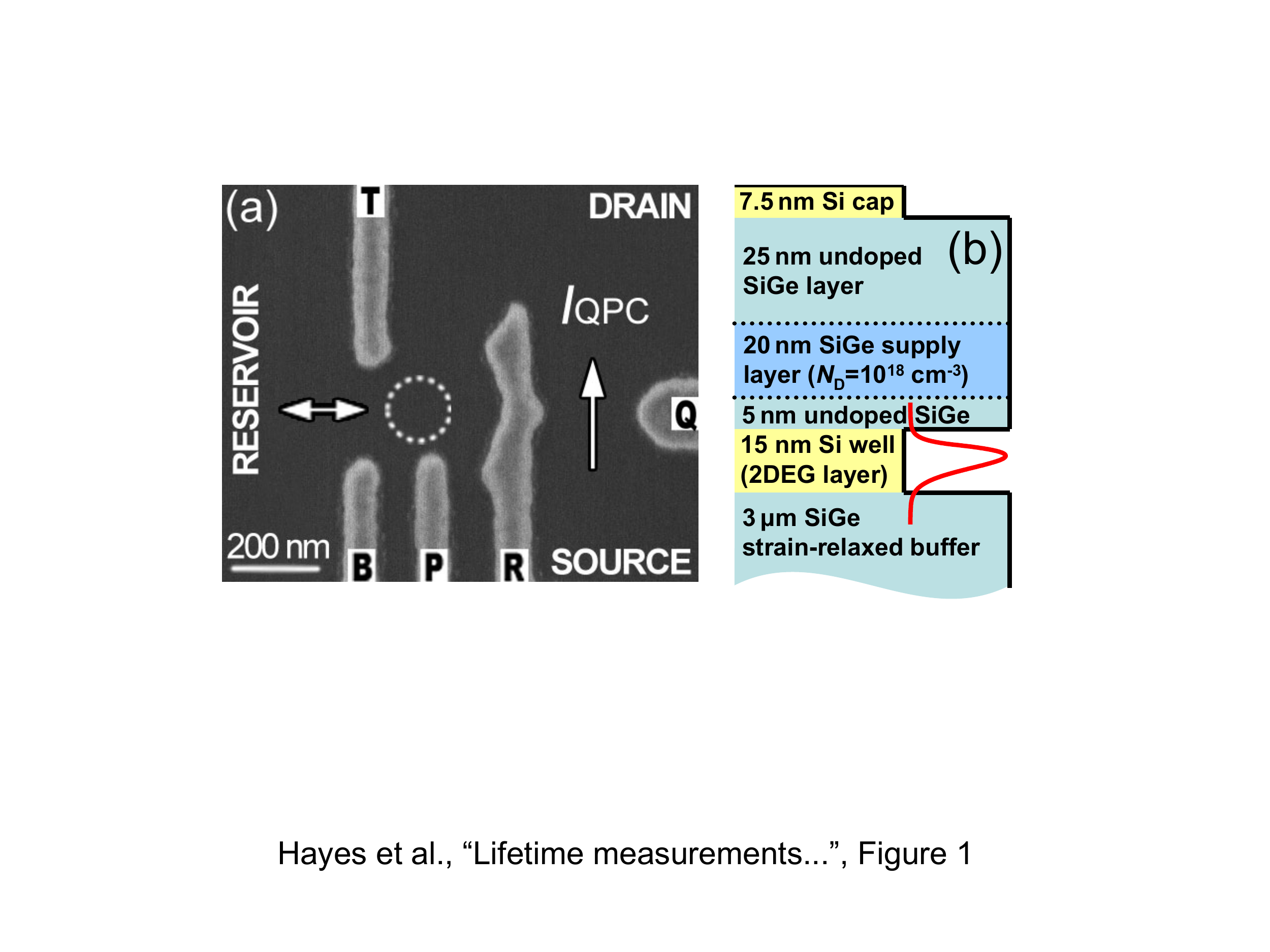}
\par\end{centering}

\caption{\label{fig:SEM-photo} (a) A SEM image of the metal electrodes on
the device surface. The dashed circle is the estimated location of
the charge disc formed by the dot electrons. The arrow shows the tunneling
channel used in all measurements. (b)~The CVD-grown $\mathrm{\textrm{Si/S\ensuremath{i_{0.7}}G\ensuremath{e_{0.3}}}}$
heterostructure with a 2DEG mobility of $10\,000$-$15\,000$~cm$^{2}$/Vs
and a charge density of $4.6$-$5.3\times10^{11}$~cm$^{-2}$ (both
measured at $4.2$~K). }

\end{figure}

\textbf{Dot Particulars }Figure \ref{fig:SEM-photo} shows the electrode
layout and epitaxial structure for our Si/Si$_{0.7}$Ge$_{0.3}$ dot.
The InGaAs/InAlAs/InP depletion dot had an almost identical electrode
geometry, but a different epitaxial structure which incorporated two
quantum wells, not one, so that the wafer could also be used for accumulation-mode
devices (described elsewhere \cite{Croke}). In our depletion-mode
devices, the upper well in the InGaAs structure played no role. 

\textbf{Measurement }The dot was electrostatically formed using conventional
techniques \cite{Elzerman}. We found, empirically, that the lowest
number of electrons on the dot, $N_{\textnormal{min}}$, can be reached
when the T--R channel is pinched off and the T--B channel kept open.
However, as the electrode voltages are lowered to drive additional
electrons off the dot, the T--B channel eventually becomes pinched
off also, preventing any further reduction in N. Although we routinely
reached $N=0$ in InGaAs dots before this happened, the lowest value
of $N$ obtained for Si/SiGe dots \textcolor{black}{was likely} $7$,
and this only after several painstaking redesigns of the electrodes
in which the gaps for the T--B and T--R channels were progressively
widened.

\begin{figure}
\begin{centering}
\includegraphics[width=3.1in]{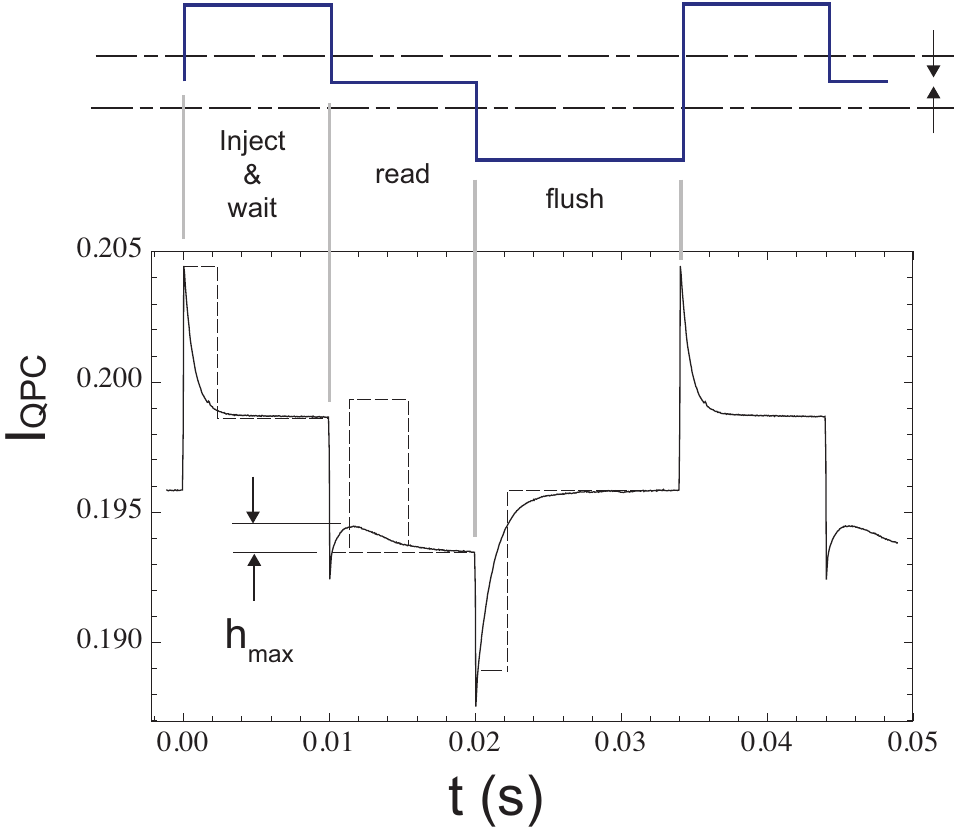}
\par\end{centering}

\caption{\label{fig:Delft-3-pulse}The Delft three-step sequence used for the
$T_{1}$ measurement \cite{Elzerman2}. The first step loads an electron
into the excited state with probability $p_{0}$, or into the ground
state with probability ($1-p_{0}$), and then waits. If the electron
is in the excited state, and survives the wait-time without decaying
to the ground state, the second step removes it from the dot and reloads
another from the reservoir into the ground state. The third step flushes
the ground state electron off the dot, so that the cycle can be repeated.
The dashed lines are an artistic rendition of instantaneous values
of $I_{\textnormal{QPC}}$ that occur during a typical cycle, the
solid curve is an actual time-average of $10\,000$ of these events. }

\end{figure}

When $N_{\textnormal{min}}$ was reached, $T_{1}$ was measured by
repeatedly applying a three-step bias sequence to one of the electrodes
(a technique first used in \cite{Elzerman2}). The response of the
charge-monitoring currrent, $I_{\textnormal{QPC}}$, to the bias changes
is twofold: (\emph{i}) a drop in the current by a fixed amount when
an electron is (spontaneously) added to the dot, and an increase when
it is removed, as shown by the dashed lines in Figure~\ref{fig:Delft-3-pulse},
and (\emph{ii}) a replication of the bias sequence due to unwanted
capacitive coupling of the driven electrode to the R--Q channel.

The key response is the rectangular pulse formed during the read interval,
which only forms when a loaded excited-state electron does \textit{not}
decay to the ground state during the wait-time. A time-average of
this rectangular pulse gives a {}``spin bump'' having the analytic
form \[
h(t)\propto p\,(e^{-t/\tau_{1}}-e^{-t/\tau_{2}})/(\tau_{2}/\tau_{1}-1)\]
where $\tau_{1}$ ($\tau_{2}$) is the tunneling time off of (onto)
the dot, and $p$ is the fraction of electrons that load into the
excited state during the inject pulse and survive the wait-time without
decaying to the ground state. For equal tunneling times this expression
reduces to $h(t)\propto p\,(t/\tau)\, e^{-t/\tau}$.

\begin{figure}
\begin{centering}
\includegraphics{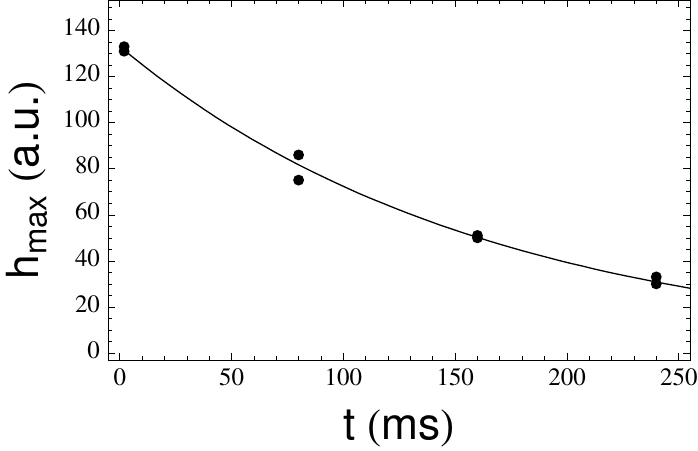}
\par\end{centering}

\caption{\label{fig:bump-height vs t}Spin bump height versus wait-time for
our Si/SiGe dot in a magnetic field of $1.5$~T. The solid curve,
an exponential fit to the data (circles), gives a $T_{1}$ of $164$~ms.}

\end{figure}

\textcolor{black}{With increasing wait-time $t_{\textnormal{wait}}$,
the fraction of instances when an unload-reload pulse is formed during
the read interval will decay exponentially as $p=p_{0\,}e^{-t_{\textnormal{wait}}/T_{1}}$
(assuming that the characteristic time $T_{1}\gg\tau$). The spin
bump height $h_{\textnormal{max}}$, plotted versus $t_{\textnormal{wait}}$,
will show the same exponential decay, as demonstrated by the data
in Fig.~\ref{fig:bump-height vs t}.}

There are two advantages of using time-averaging rather than the pulse-counting
of \cite{Elzerman2}: (\emph{i}) the high signal-to-noise ratio required
to reliably detect individual pulses is not needed, and (\emph{ii})
the false counts due to random telegraph signals (RTS) that will unavoidably
occur when the read level is near the lower Zeeman level contribute
trivially, because the average of RTS-induced pulses is a straight
line. What is lost by using the averaging approach is a direct measure
of the branching ratio, $p_{0}$. However, $p_{0}$ is not needed
for a determination of $T_{1}$.

\textbf{Zeeman splitting }The spin-bump appears when, and only when,
the amplitude of the inject-and-wait pulse is sufficient to reach
the first excited state. This suggests a step-and-reach technique
for measuring the Zeeman splitting. One places the read level a known
amount above the ground state (the \textit{step} process), and then
progressively \textit{reaches} up with the inject pulse until a bump
appears. The exact point of appearance of the bump is determined experimentally
by measuring the bump height for several values of step-plus-reach
amplitude just beyond the first appearance of the bump, and then extrapolating
backward to determine the zero-intercept. To convert the combined
step and reach amplitudes to an actual energy, we used a temperature
sweep and the known relation between the width of a Coulomb Blockade
peak and the dot\textquoteright{}s temperature \cite{Meirav} to determine
both the lever arm, $\alpha$, and the effective base temperature
of the system, $12.55$~$\textrm{V}_{\textnormal{Plunger}}$/eV and
$48$~mK, respectively. 

\begin{figure}
\begin{centering}
\includegraphics{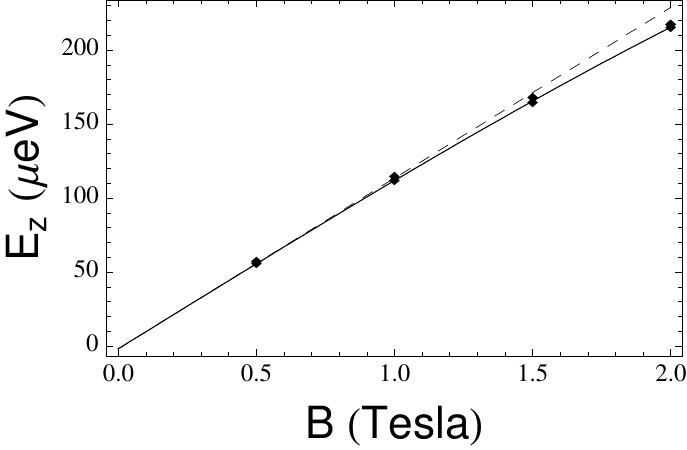}
\par\end{centering}

\caption{\label{fig:Zeeman-splitting}The Zeeman splitting as a function of
magnetic field. The solid curve is a fit with $E_{\textnormal{Z}}=a+bB+cB^{3}$,
the dashed curve is the linear portion of that expression.}

\end{figure}

The Zeeman splitting was measured for several values of the magnetic
field using this technique. Two sets of measurements were made, one
with the magnetic field increasing with each successive point, the
other with it decreasing, in order to look for hysteretic effects
(which we did not see). These data are plotted in Fig.~\ref{fig:Zeeman-splitting}.
One sees immediately that there is a small but non-negligible deviation
from a straight line at higher values of $B$, which we suspect may
be caused by a small component of the magnetic field normal to the
2DEG. We have fit the $9$ measured points to the somewhat arbitrary
function $E_{\textnormal{Z}}=a+bB+cB^{n}$, using values of $n$ ranging
from $2$ to infinity, and have found that these fits give a zero-field
value of $g$ that ranges from $2.1$ to $1.9$. Although the $g$-value
that results from the fit having the lowest variance and proper symmetry
($n=3$) is $1.99$, all fits appear equally acceptable to the eye,
making it difficult to choose one over another. Our estimate of $g$
is thus $2.0\pm0.1$, in agreement with theoretical expectations and
measurements on impurity-bound electrons in Si \cite{T1}. 

\textbf{Multi-electron dots }Our measurements of $T_{1}$ for Si/SiGe
are for an $N$-electron dot that we believe \textcolor{black}{had
$7$ electrons ($5$ to $7$ electrons} if based exclusively on the
energy spacings between Coulomb Blockade peaks, or, more definitively,
$7$ electrons when further constrained by the filling sequence of
spin states as discussed below). Measurements of $T_{1}$ at lower
values of $N$ were not possible because the dot {}``closed''. 

To see what would happen at values of $N>1$ in dots that can reach
$N=0$, we looked for spin bumps on our InGaAs dot, starting with
$N=1$, and progressing all the way up to $N=9$. We found a spin
bump at $N=1,\,3,\,5,$ and $7$, but none for $9$, and none for
$N=2,\,4,\,6,$ and $8$. These sightings are consistent with Pauli
filling of an asymmetric dot \cite{Reimann}, in which the spin values
would be $\nicefrac{1}{2},\,0,\,\nicefrac{1}{2},\,0\ldots$ for $N=1,\,2,\,3,\,4\ldots$
electrons. The measured value of $T_{1}$ for $N=7$ was $12.8$~ms,
which is shorter than the $41.5$~ms measured for $N=1$, but not
dramatically so. 

For Si/SiGe dots, however, which have an extra degeneracy due to valleys,
there is reason to believe (e. g., see \cite{Hada}) that the filling
sequence for an asymmetric dot should be $\nicefrac{1}{2},\,1,\,\nicefrac{1}{2},\,0,\,\nicefrac{1}{2},\,1,\nicefrac{1}{2},0\ldots$
when the Zeeman energy exceeds the valley splitting. For our lowest
$N$-value, the measured Zeeman splitting and $g$-value of $2.0$
imply a spin of $\nicefrac{1}{2}$ (or greater). If the spin is $\nicefrac{1}{2}$,
the next higher value of $N$ should have an $s=0$ or $s=1$ ground
state. Measurements of the $N+1$ dot showed that the ground state
had $s=0$, and the excited state $s=1$, with a surprisingly-small
singlet-triplet splitting of approximately $36$~$\mu$eV \cite{to be published}.
Thus we contend that our measured values of $T_{1}$ are for an unpaired
spin of $\nicefrac{1}{2}$, and are representative of (although probably
somewhat shorter than) the value that would be obtained for a single-electron
spin-$\nicefrac{1}{2}$ dot. 

\textbf{Theory} At reasonably strong spatial confinement, low temperatures,
and small-to-moderate magnetic fields, the single-phonon admixture
mechanism dominates the \emph{intrinsic} spin flip ($T_{1}$ relaxation
process) of zero-dimensional conduction electrons in III-V materials
\cite{Khaetskii}. It can be understood as follows. Spin-orbit (SO)
coupling\emph{ }admixes\emph{ }higher orbital states with opposite
spin projections into the eigenstates of the lowest spin doublet,
thus allowing, in principle, intra-doublet transitions (i.~e., spin
flip) to be induced by strictly spin-independent interactions with
the environment, e.~g., with lattice vibrations. Exact cancellation
in the transition matrix element, enforced by the time-reversal symmetry
of the total electron Hamiltonian (including the SO admixing terms),
should be broken by a magnetic field to yield a finite spin relaxation
rate. 

At low magnetic fields the Zeeman splitting of the SO-mixed doublet
(and, thus, the required energy transfer to the lattice phonon) is
small, so that, when present, piezoelectric electron-phonon coupling
dominates, resulting, at temperatures $T\ll g\mu_{B}B/k_{B}$, in
a well-known \cite{Khaetskii,Woods,Stano} functional dependence \[
T_{1\textnormal{P}}^{-1}=P\eta^{2}B^{5}/E_{0}^{4}\]
obtained in the (valild for our case) approximation of long-wavelength
resonant phonons. $E_{0}$ is the spatial quantization energy. SO
coupling is assumed to be dominated by linear-in-$k$ SO terms in
the electron dispersion; their strength is quantified by a material-
and structure-specific constant $\eta$, which is, in general, anisotropic.
In non-piezoelectric materials, but also in piezoelectric materials
at larger magnetic fields, energy transfer to phonons is facilitated
by a deformation potential coupling resulting (e.~g., \cite{Stano})
in \[
T_{1\textnormal{D}}^{-1}=D\eta^{2}B^{7}/E_{0}^{4}\]
at low temperatures. At temperatures $T\gg g\mu_{B}B/k_{B}$, the
spin-flip rate is given by $T_{1\textnormal{P},\textnormal{D}}^{-1}\, k_{B}T/g\mu_{B}B$. 

\begin{figure}
\begin{centering}
\includegraphics[bb=30bp 20bp 700bp 560bp,clip,width=3.1in]{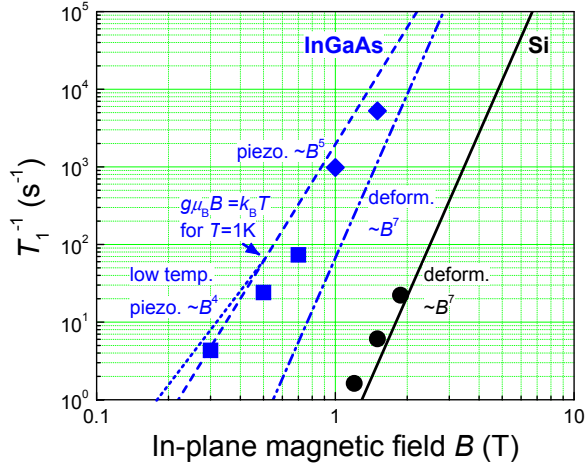}
\par\end{centering}

\caption{\label{fig:Inverse-lifetimes}Inverse spin lifetimes as a function
of magnetic field. Circles are for the Si depletion dot, squares (diamonds)
for the InGaAs depletion (accummulation) dot. $B$ is always parallel
to the 2DEG \textcolor{black}{(along the $[010]$ axis).} The straight
lines are theoretical predictions for the two material systems.}

\end{figure}

Moderate electric fields $\lesssim2$~mV/nm are typically present
in our dots. In InGaAs structures, a bulk-inversion-asymmetry-induced
(BIA or Dresselhaus) SO term with $\beta\sim1\,600$~$\mu$eV~nm
dominates the structure-induced (SIA or Rashba) term ($\lesssim500$~$\mu$eV~nm),
making $\eta\approx\beta$ a field-independent and \emph{isotropic}
constant, so that no in-plane anisotropy is to be expected for $T_{1}$
in a circular dot. $P$ and $D$ are material-, but not structure-specific
parameters (apart from a weak $g$ factor dependence). When expressing
$T_{1}^{-1}$ in inverse seconds, $\eta$ in $\mu$eV~nm, the \emph{in-plane}
$B$ in Tesla, and $E_{0}$ in meV, the InGaAs-specific constants
$P$ and $D$ are found to have numeric values of $7.6$ and $0.26$,
respectively \cite{derived}. These low-temperature asymptotics of
$T_{1\textnormal{P}}^{-1}$ and $T_{1\textnormal{D}}^{-1}$ are shown
in Fig.~\ref{fig:Inverse-lifetimes} by dashed and dash-dotted lines,
respectively, for the relevant $E_{0}=10$~meV. The dotted line is
the high temperature $B^{4}$-asymptote of $T_{1\textnormal{P}}^{-1}$
calculated for $T=1$~K. Our measurement temperatures are always
much lower.

The Si crystal symmetry allows neither piezoelectricity nor a BIA
SO term. SIA is dwarfed by an especially weak bulk SO coupling for
$\Delta$ valleys (estimated to be only $3$--$4$~$\mu$eV~nm in
comparable structures \cite{SIA in Si}). It was suggested in \cite{Nestoklon}
that linear-in-$k$ SO terms in (001) Si heterostructures are dominated
by contributions of a low-symmetry heterointerface. For our typical
electric fields, we estimate, with the help of Fig.~3 in \cite{Nestoklon},
that $\eta$ has a maximum value of $20$--$30$~$\mu$eV~nm for
a perfect interface, but a much lower value in the presence of interface
imperfections.\textcolor{red}{ }In Fig.~\ref{fig:Inverse-lifetimes},
$T_{1\textnormal{D}}^{-1}$ in Si (with $D=4.3\times10^{-3}$) is
shown by solid lines for (the numerically simulated) $E_{0}=2$~meV.

Other spin-lattice relaxation mechanisms were proposed for Si dots,
in particular: (\emph{i}) the modulation of the hyperfine coupling
by phonon deformation, which has a $B^{3}$ field dependence \cite{Khaetskii,Khaetskii2},
and (\emph{ii}) the phonon-induced modulation of the \emph{bulk} electron
$g$-factor, with a $B^{5}$-dependence and a strong sensitivity to
the orientation of the in-plane magnetic field with respect to the
main crystallographic axes \cite{Glavin}. As formulated, these mechanisms
are not directly related to the heterointerface properties. They should
exceed $T_{1\textnormal{D}}^{-1}$ at extremely small magnetic fields,
but the estimated times are too slow to be observed in our experiments.

\textbf{Acknowledgements} We gratefully acknowledge helpful discussions
with M.A. Eriksson, H.W. Jiang, and C.M. Marcus. This work was sponsored
by the United States Department of Defense. The views and conclusions
contained in this document are those of the authors and should not
be interpreted as representing the official policies, either expressly
or implied, of the United States Department of Defense or the U.S.
Government.

\end{document}